\pdfoutput=1

\documentclass[8.5pt,twoside,twocolumn]{article}
\usepackage[super,sort&compress,comma]{natbib}
\usepackage{mhchem}
\usepackage{times,mathptm}
\usepackage{sectsty}
\usepackage{balance}

\usepackage{graphicx} 
\usepackage{lastpage}
\usepackage[format=plain,justification=raggedright,singlelinecheck=false,font=small,labelfont=bf,labelsep=space]{caption}
\usepackage{fancyhdr}
\begin{document}

\twocolumn[
 \begin{@twocolumnfalse}
\noindent\LARGE{\textbf{Feature-rich electronic properties of aluminum-doped graphenes}}
\vspace{0.6cm}

\noindent\large{\textbf{Shih-Yang Lin,\textit{$^{a}$} Yu-Tsung Lin,\textit{$^{a\dag}$} Ngoc Thanh Thuy Tran,\textit{$^{a}$} Wu-Pei Su,\textit{$^{b\ddag}$} Ming-Fa Lin,\textit{$^{a\ast}$}}}\vspace{0.5cm}

\noindent\textit{\small{\textbf{Received Xth XXXXXXXXXX 20XX,
Accepted Xth XXXXXXXXX 20XX\newline First published on the web Xth
XXXXXXXXXX 200X}}}

\noindent \textbf{\small{DOI: 10.1039/b000000x}}
\vspace{0.6cm}

\noindent \normalsize{The electronic properties of aluminum-doped graphenes enriched by multi-orbital hybridizations are investigated using first-principles calculations. The feature-rich electronic structures exhibit the quasi-rigid red shifts of the carbon-created energy bands, the Al-dominated valence and conduction bands, and many free electrons in the conduction Dirac cone. These are directly reflected in the special structures of density of states (DOS). The Al- and alkali-induced high free carrier densities are almost the same. There exist certain important differences among the Al-, alkali- and halogen-doped grapehenes, such as, the buckled or planar graphene, the preserved or seriously distorted Dirac cone, the existence of the adatom-dominated valence bands, the free electrons or holes, the degenerate or splitting spin-related states, and the simple or complicated peaks in DOS. The similarities and differences mainly come from the diverse orbital hybridizations in adatom-C bonds, as indicated from the atom-dominated bands, the spatial charge distributions and the orbital-projected DOS.} \vspace{0.5cm}
\end{@twocolumnfalse}
  ]
\

\section{INTRODUCTION}
\footnotetext{\textit{$^{a}$~Department of Physics, National Cheng Kung University, 701 Tainan, Taiwan.}}
\footnotetext{\textit{$^{b}$~Department of Physics, University of Houston, 77054 houston, TX.}}
\footnotetext{\dag~E-mail: xavier.rusta@gmail.com}
\footnotetext{\ddag~E-mail: wpsu@uh.edu}
\footnotetext{$\ast~$ E-mail: mflin@mail.ncku.edu.tw}

Graphene, with nanoscale thickness and honeycomb lattice, has attracted a lot of theoretical and experimental researches.\cite{novoselov2005two,schedin2007detection,geim2009graphene,lee2008measurement,berger2006electronic,balandin2008superior,liu2011chemical,sutter2009electronic,zhong2012stacking,lai2008magnetoelectronic,pereira2009strain} It exhibits many unusual properties, such as, the tunable carrier densities,\cite{lee2008measurement} the ultrahigh carrier mobility,\cite{berger2006electronic} and the Dirac-cone structure.\cite{balandin2008superior}
The graphene-based nano-materials might have high potentials in the next-generation devices, e.g., the ultrafast rechargeable aluminium-ion battery,\cite{lin2015ultrafast,rani2013fluorinated} the high-frequency transistors,\cite{lin2010100,schwierz2010graphene} and the large reversible lithium storages.\cite{paek2008enhanced,wang2009graphene}
Monolayer graphene is a zero-gap semiconductor with vanishing density of states (DOS) at the Fermi level ($E_{F}$), while few-layer graphenes are semimetals with low free carrier density.
The greatly enhanced free-carrier density or the gap opening is an important issue for the potential applications.
The electronic properties can be easily modulated by the adatom dopings,\cite{liu2011chemical,wei2009synthesis,joucken2015charge} layer numbers,\cite{sutter2009electronic,lauffer2008atomic} stacking configurations,\cite{zhong2012stacking,tran2015configuration} electric and magnetic fields,\cite{lai2008magnetoelectronic,lin2015magneto} and mechanical strains.\cite{pereira2009strain,wong2012strain}
Among these methods, atom adsorptions is the most effective way to dramatically change the energy gaps or induce the metal-semiconductor transitions.
Specifically, Al atoms in the aluminium-ion battery can largely enhance the current densities with low cost and flammability.
This work is focused on the unusual electronic properties created by the aluminum adsorption on graphene. By the detailed calculations, the critical orbital hybridizations are proposed to comprehend the essential physical properties. A thorough comparison among the Al-, alkali-, and halogen-doped graphenes is also made.

To date, the Al-related graphene systems are one of the most widely studied materials, since the high sensitivity of Al atoms can be applied to environment and energy engineering.
When a certain C atom in hexagon is replaced by Al, this system can serve as a toxic gas sensor, according to the theoretical predictions.\cite{chi2009adsorption,ao2008enhancement}
The formaldehyde (H$_{2}$CO) sensor is attributed to the ionic bonds associated with the charge transfer and the covalent bonds due to the overlap of orbitals.\cite{chi2009adsorption} Carbon monoxide can be detected by the drastical change in the electrical conductivity before and after molecule adsorption.\cite{ao2008enhancement} Also, this Al-related system can serve as a potential hydrogen storage material at room temperature, in which the functional capacity is largely enhanced by the Al atoms.\cite{ao2009doped} The similar functionality is predicted to be revealed in Al-doped graphenes.\cite{ao2010high}
Most importantly, Al atoms play an critical role in the dramatic enhancement of current densities in aluminium-ion battery, where the predominant AlCl$_{4}^{-}$ molecules are intercalated and de-intercalated between graphite layers during charge and discharge reactions, respectively.\cite{lin2015ultrafast} The diverse physical and chemical phenomena mainly come from the distinct orbital bondings. The multi-orbital hybridizations are expected to dominate the essential properties of the Al-doped graphenes, e.g., the atom-related energy bands, the spatial charge distributions and the orbital-projected special structures in DOS.

The interactions between graphene and alkali-metal atoms has received considerable attention in recent years,\cite{chan2008first,jin2010crossover,praveen2015adsorption} mainly owing to the efficient n-type doping. Specifically, the Dirac-cone structures in the alkali-intercalated\cite{virojanadara2010epitaxial,sugawara2011fabrication,gruneis2008tunable} or -doped\cite{papagno2011large} exhibits an obvious red shift. The strong evidence of high free electron density is confirmed by angle-resolved photoemission spectroscopy (ARPES).\cite{virojanadara2010epitaxial,sugawara2011fabrication,gruneis2008tunable,papagno2011large} On the other hand, the p-type doping can be realized in the halogen-related graphenes.\cite{walter2011highly,vinogradov2012controllable,wang2014heteroatom}
The high hole concentration is created by the p-dopants, F, Cl, and Br.\cite{wang2014heteroatom} Moreover, the ARPES measurements have verified the blue shift of the Dirac-cone structure in the F- and Cl-intercalated graphene.\cite{walter2011highly,vinogradov2012controllable} Although some theoretical studies on halogen-doped graphene have been made for the specific adatoms or applications,\cite{leenaerts2010first,samarakoon2011structural,sahin2012chlorine,zbovril2010graphene} a comprehensive and comparative work is required for the orbital bonding-enriched phenomena, e.g., the similarities and differences in energy bands, free carriers and spin configurations.

In this paper, the geometric and electronic properties of the Al-doped graphenes are investigated by using the first-principles calculations. The dependence on the concentration and arrangement of Al adatoms is explored in detail; furthermore, the relations among the Al-, alkali- and halogen-doped graphenes are
discussed extensively. Bond lengths, heights, absorption energies, carbon- or adatom-dominated energy bands, band-edge states, free carrier density, charge distribution, spin configuration and density of states are included in the calculations. The spatial charge distribution and the orbital-projected DOS are useful to comprehend how the electronic properties are enriched by the orbital hybridizations in the carbon-adatom bonds. Apparently, this work shows the diversified electronic properties, including the rigid shift or the distortion of the Dirac-cone structure, the dramatic changes in the $\pi$ and $\pi^{*}$ bands, the metallic behaviors due to free electrons or holes, the creation of the adatom-dominated valence and conduction bands, and the degeneracy or splitting of the spin-related energy bands. Such properties are strongly affected by the distinct adatom adsorptions. They are further reflected in a lot of special structures of DOS. The predicted energy bands and DOS could be verified by ARPES\cite{zhou2008metal,schulte2013bandgap} and scanning tunneling spectroscopy (STS),\cite{tapaszto2008tuning,gyamfi2011fe} respectively.

\section{COMPUTATIONAL DETAILS}
For the geometric and electronic properties of the Al-absorbed graphene systems, the first-principle density functional theory calculations are performed by using the Vienna \emph{ab initio} simulation package.\cite{kresse1996efficient} The exchange-correlation energy of interacting electrons is calculated within the Perdew-Burke-Ernzerhof functional (PBE) \cite{perdew1996generalized} using the generalized gradient approximation. The projector-augmented wave pseudopotentials \cite{blochl1994projector} are employed to characterize the electron-ion interactions. The wave functions are built from the plane waves with a maximum energy cutoff of $500$ eV. The vacuum distance along z-axis is set to be 15 {\AA} for avoiding the interaction between adjacent cells. The first Brillouin zone is sampled in a Gamma scheme along the two-dimensional periodic direction by $12\times12\times1$ k points for all structure relaxations, and by $100\times 100\times1$ for further studies on electronic properties. The convergence criterion for one full relaxation is determined by setting the Hellmann-Feynman forces weaker than $0.01$ eV/{\AA} and the total energy difference of $\Delta\,E_t<10^{-5}$ eV.

\section{RESULTS AND DISCUSSION}

Six typical supercells, namely $2\times2$ $(12.5 \%)$, $3\times3$ $(5.6 \%)$, $2\sqrt{3}\times2\sqrt{3} $ $(4.2 \%)$, $4\times4 $ $(3.2 \%)$, $5\times5 $ $(2.0 \%)$, and $3\sqrt{3}\times3\sqrt{3}$ $(1.9 \%)$, are chosen to study the Al-doped effects. The optimal adatom position is the hollow site, regardless of Al-concentration and distribution. As the highest Al-concentration is reduced to the lowest one, the carbon-carbon bond lengths nearest to adatoms are changed from $1.441$ {\AA} to $1.419$ {\AA}, but the others from $1.431$ {\AA} to 1.413 {\AA} (Table 1). The lattice constant is expanded only $1.6 \%$ for the $2\times2$ case, compared with pristine graphene. The Al-C bond length and the height (h) of adatom are, respectively, $2.54$ {\AA} and $2.00$ {\AA}; they hardly depend on the Al concentration. Such bond lengths are related to the orbital hybridizations in the Al-C bonds. For the alkali- and halogen-doped systems, a $4\times4$ supercell is used to investigate the adatom effects (Table 2). (Li,Na,K) are most stable at the hollow sites with h$= 1.70$, $2.18$; $2.52$ {\AA}, respectively. On the other hand, the top sites are the optimal position for (F,Cl,Br), accompanied with h$= 1.56$, $2.91;$ $3.13$ {\AA}. Specifically, the F-doped graphene has the lowest adsorption energy and a buckled structure (h$=0.40$ {\AA} for the nearest C; Fig. 4(e)), indicating the non-negligible $sp^3$ bonding on graphene layer. The much difference in height will greatly diversify the electronic properties of hallogen-doped graphene.

\begin{figure}[h]
\centering
  \includegraphics[width=1.0\linewidth]{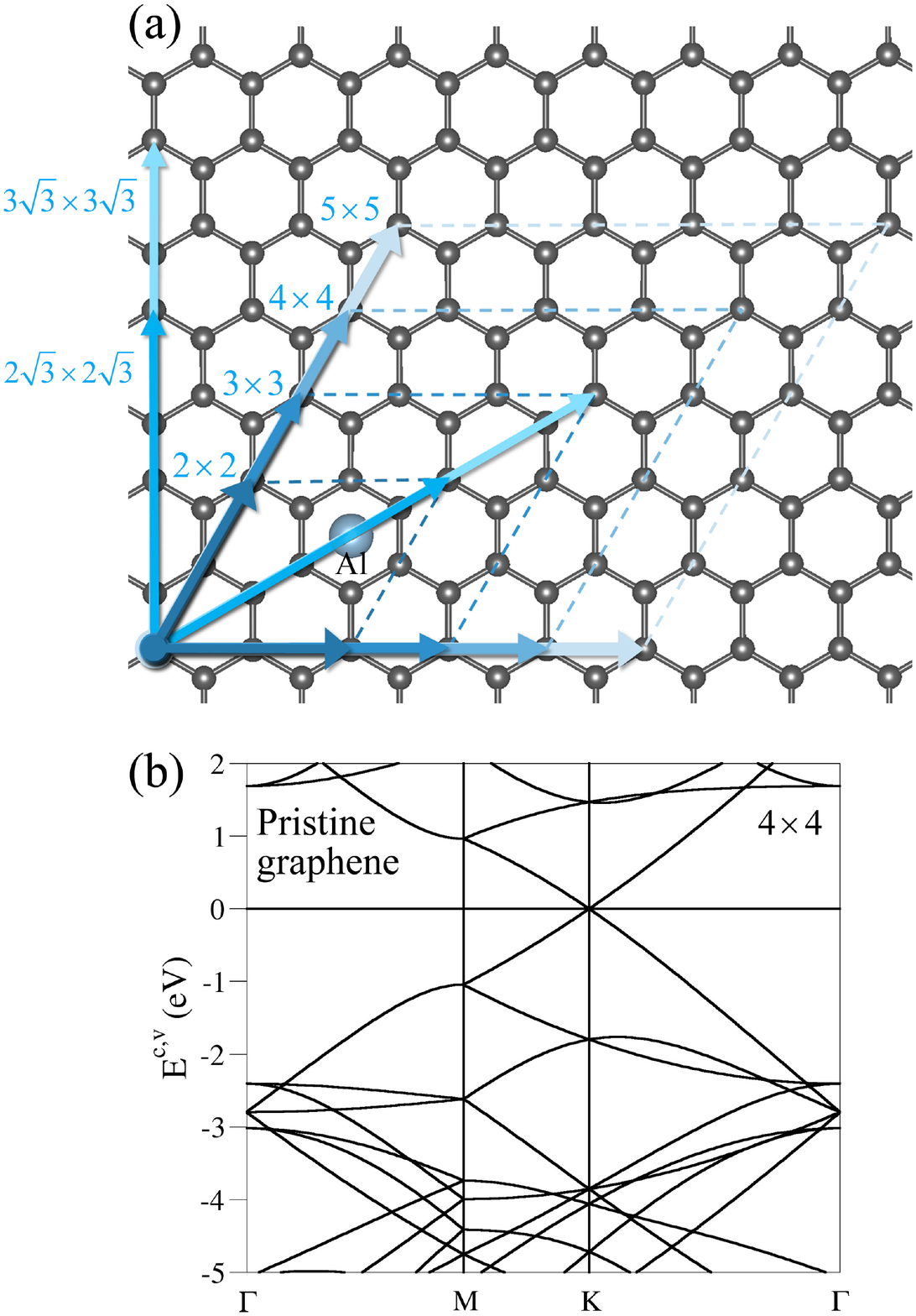}
  \caption{(a) Geometric structures for various Al-doped graphenes, and (b) band structure of pristine graphene in a enlarged $4\times\,4$ cell.}
  \label{fgr:1}
\end{figure}

\begin{table}
\small
\caption{Bond lengths and heights for various Al-doped graphenes.}
\label{t1}
  \begin{tabular*}{0.5\textwidth}{@{\extracolsep{\fill}}cccccc}
\hline
& \multicolumn{3}{c}{ Bond length ({\AA})} & & \\
Unit cell&Nearest&Next nea-&Al-C&Lattice& Al \\
 &C-C&rest C-C&&expansion&height({\AA})\\
\hline
$2\times2$ & 1.441 & 1.431 &2.544& 1.63 \% & 2.097\\
$3\times3$ & 1.440 & 1.431 &2.538& 1.36 \% & 2.090\\
$2\sqrt{3}\times2\sqrt{3}$ &2.543& 1.440 & 1.428 & 1.17 \% & 2.096\\
$4\times4$ & 1.436 & 1.428 &2.544& 1.12 \% & 2.110\\
$5\times5$ & 1.419 & 1.413 &2.571& 1.06 \% & 2.128 \\
$3\sqrt{3}\times3\sqrt{3}$ &2.563& 1.438 & 1.428 & 1.02 \% & 2.110\\
\hline
\end{tabular*}
\end{table}

\begin{table}
\small
\caption{Adsorption energies, heights and bond lengths for various adatom-doped graphenes in a $4\times4$ unit cell.}
\label{t1}
  \begin{tabular*}{0.5\textwidth}{@{\extracolsep{\fill}}ccccc}
\hline
Adatom & Energy & Lattice & Adatom& Adatom-C \\
 &  (eV) & expansion & height ({\AA}) & bonds ({\AA})\\
\hline
Al & -1.010 & 1.12\% & 2.11 & 2.554\\
Li & -1.207 & 0.94\% & 1.70 & 2.241\\
Na & -0.414 & 1.04\% & 2.18 & 2.617\\
K & -0.750 & 1.13\% & 2.52 & 2.937\\
F & -3.099 & 0.84\% & 1.56 & 1.566\\
Cl & -0.879 & 0.82\% & 2.91 & 2.912\\
Br & -0.798 & 1.02\% & 3.13 & 3.221\\
\hline
\end{tabular*}
\end{table}

The electronic structures are enriched by tuning the Al concentration, specially for the free electrons in the conduction Dirac-cone.
Pristine graphene has an isotropic Dirac-cone structure within $|E^{c,v}|\le\, 1$ eV, in which the intersecting points are just located at the K point, as shown in Fig. 1(b).
The $\pi$-electronic linear valence and conduction bands are formed by the $2p_z$ orbitals.
As the state energy increases, they gradually become parabolic energy dispersions, and there are special saddle points at the $\Gamma$ point with energies $-2.4$ eV and $1.7$ eV.
Also, the $\sigma$-electronic energy bands, with a local maximum of $E^{v}= -3.0$ eV at the $\Gamma$ point, mainly come from $2p_x$ and $2p_y$ orbitals.
Energy bands are dramatically changed by the Al-adatom absorption.
The Dirac-cone structure is almost preserved in the Al-absorbed graphene; furthermore, it has a rigid red shift, as indicated in Figs. 2(a)-2(f).
The Fermi level is located at the conduction cone, indicating a great amount of Al-induced free electrons.
The Dirac points might be merged together or slightly separated.
Their energies gradually grow with the decreasing Al concentration, but almost remain at $\sim-1$ eV for $D_{Al}\, \textless$ $5.5\%$.
It is also noticed that the similar red shift is revealed in the $\sigma$ energy bands.
Specially, Al atoms make much contribution to certain energy bands with weak dispersions in the range of $-3.5$ eV $\le\,E^{c,v}\le\,-2.5$ eV and $0.3$ eV $\le\,E^{c,v}\le\,1.5$ eV, where the amount of contribution indicated by the radius of green circles. In addition, the Al-dominated valence and conduction bands, respectively, come from the $3s$ and ($3p_x$,$3p_y$) orbitals (details in Fig. 5).

\begin{figure}[h]
\centering
  \includegraphics[width=1.0\linewidth]{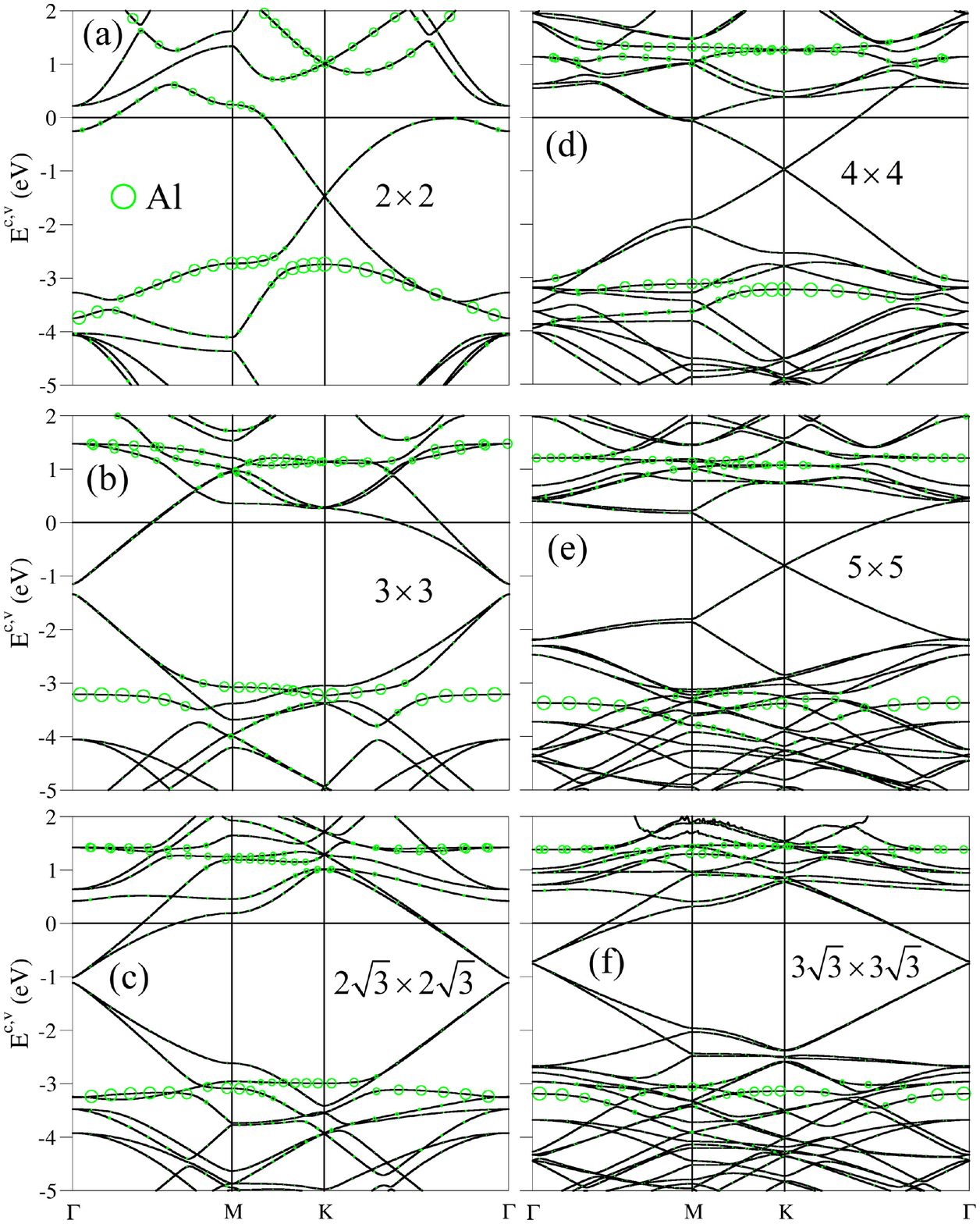}
  \caption{Band structures of Al-doped graphene for various adatom adsorptions: (a) $2\times\,2,$ (b) $3\times\,3,$ (c) $2\sqrt 3\times\,2 \sqrt 3\,$ (d) $4\times\,4,$  (e) $5\times\,5,$ and (f) $3\sqrt 3\times\,3 \sqrt 3\,$.}
  \label{fgr:2}
\end{figure}

The alkali-metal atoms exhibit the similar doping effects, as revealed by the Al atoms. The Li-, Na-, and K-adsorbed systems, with a $4\times4$ supercell, present the almost same band structure within $-3$ eV $\le\,E^{c,v}\le\,0$, such as the Fermi-momentum states, the low-lying linear dispersions, and the Dirac-point energy $E_{D}$, as shown in Figs. 3(a)-3(c). These features are almost identical to those of the Al-doped graphene (Fig. 2(e)). That is to say, such n-doped systems possess the same Dirac-cone structure and free carrier density. Specially, there exists a low-lying conduction band with a weak dispersion in $0.2$ eV $\le\,E^{c,v}\le\,0.9$ eV, being mostly contributed by the alkali adatoms (green circles in Figs. 3(a)-3(c)).
However, the adatom-dominated band structures in Al-doped graphene have an unoccupied conduction band ($|E^{c}| \sim 1.2$ eV) and an effective valence band ($|E^{v}| \sim -3.2$ eV) (Fig. 2(e)). The former is similar to that of alkali-doped graphenes. Furthermore, the latter originates from certain orbital hybridizations in the C-Al bond, being absent in alkali-doped graphenes within $-5$ eV $\le\,E^{c,v}\le\,0$. The Al-dominated valence states are occupied by two electrons of each adatom, and the remaining one electron is distributed to the conduction Dirac-cone structure according to the electronic state energies. The same free carrier density is also revealed in the alkali-doped graphenes. These can account for the almost identical Dirac-cone structures in such n-doped systems. The Al- and alkali-doped can create very high free carrier density, so they are expected to have high potentials in the next-generation high-capacity batteries\cite{lin2015ultrafast,rani2013fluorinated} and energy storage.\cite{ao2009doped,ao2010high}

\begin{figure}[h]
\centering
  \includegraphics[width=1.0\linewidth]{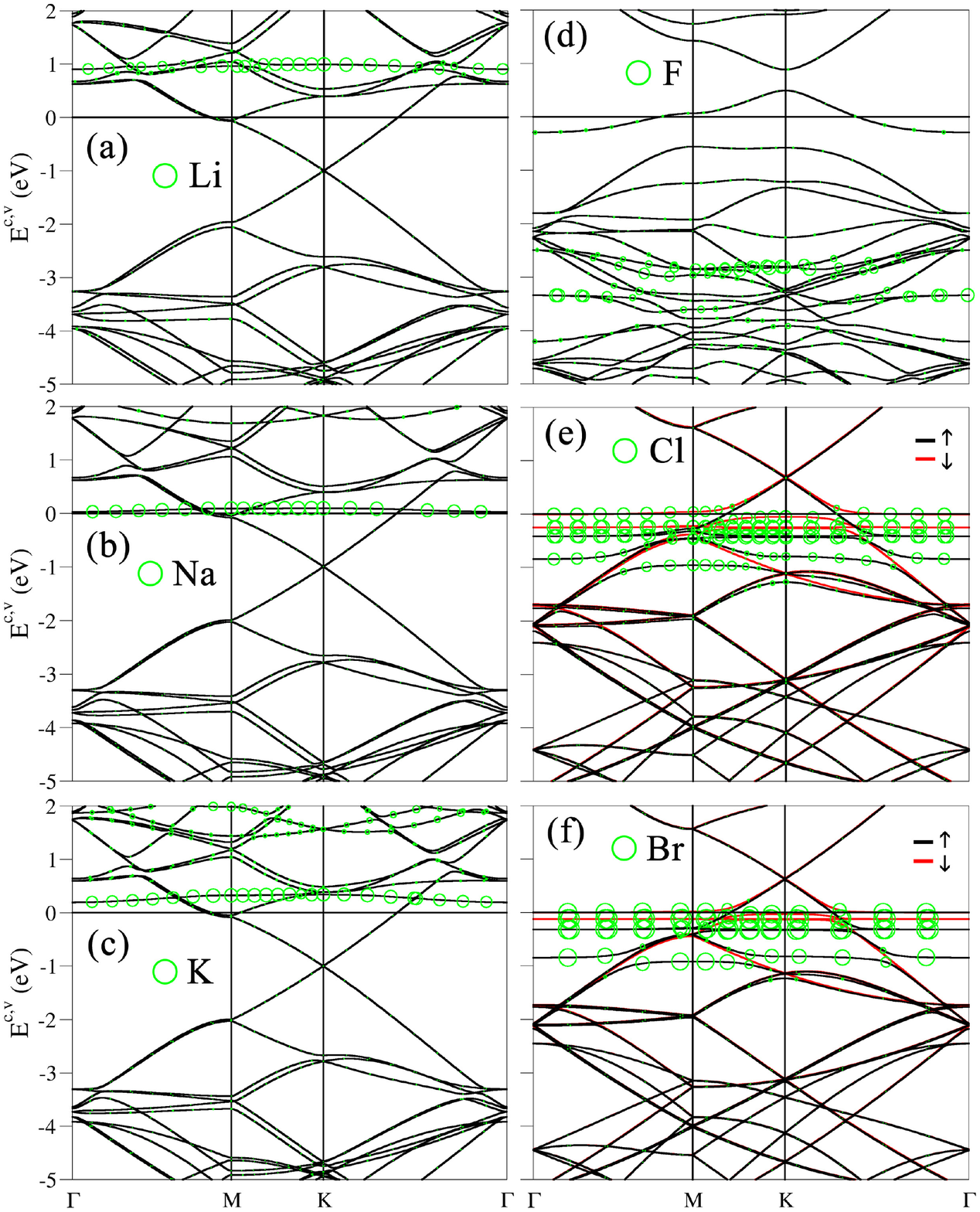}
  \caption{Energy bands of alkali- and halogen-doped graphenes in a $4\times\,4$ unit cell for (a) Li, (b) Na, (c) K, (d) F, (e) Cl, and (f) Br. The last two systems have the spin-split bands.}
  \label{fgr:3}
\end{figure}

Free carriers between $E_{D}$ and $E_{F}$ deserve a closer examination. The free electron density ($\sigma_f$) can be directly evaluated from the enclosed area in the 2D wave-vector-space. $\sigma_f$ is $\sim 1.20\times10^{14}$ cm$^{-2}$ in the $4\times4$ case, being consistent with one electron contributed by each adatom per unit cell. This clearly illustrates that $\sigma_f$ is proportional to the adatom-concentration. Under such cases, the alkali-doped graphenes also have the same free electron density distributed in the conduction Dirac-cone structure. When the adatom concentration is sufficiently high ($\ge\,12.5 \%$), the alkali-dominated conduction band-edge states are lower than the Fermi level and even overlap with the Dirac cone. Specifically, the bottoms of the Li-, Na-, and K-dominated parabolic bands are located at $E^{v}=-0.1$ eV, $-0.4$ eV, and $-0.5$ eV for the $2\times\,2$ case, respectively. However, one electron contributed by adatom is mostly distributed in conduction cone and partly in adatom-dominated band. That is to say, the adatom concentrations in metallic graphene systems dominate the free carrier density and the carrier distribution in energy bands.

The halogen-doped graphenes are in sharp contrast to the Al- and alkali-doped systems in their electronic properties, such as the Dirac-cone structure, the Fermi level, the free carrier density, the adatom-dominated bands, and the spin configurations.
The Dirac-cone structure is destroyed or slightly distorted, depending on the chemical bondings between halogen and graphene, as shown for F-, Cl-, and Br-doped graphenes in Figs. 3(d)-3(f).
The Fermi level is located at the valence Dirac cone, i.e., free holes appear between $E_{D}$ and $E_{F}$. The free carrier density is lower than that in Al-doped graphene.
The linearly intersecting bands become parabolic bands with a separated Dirac point for the F-doped system (Fig. 3(d)), while the Cl- and Br-doped systems only present the slight distortions (Figs. 3(e) and 3(f)).
This clearly reflects the more complicated orbital hybridizations in F-C bonds, being attributed to the lower height of adatom. Such hybridizations lead to the drastic changes in the $\pi$ and $\sigma$ bands; that is, the quasi-rigid blue shifts of the pristine energy bands are absent. Moreover, the strength of chemical bonds can determine where the adatom-dominated bands are located. The rather strong F-C bonds cause the adatom-dominated valence bands to be situated in the deeper energy range of $-3.5$ eV $\le\,E^{v}\le\,-2.5$ eV.
On the other hand, the Cl- and Br-dominated energy bands appear in $-1$ eV $\le\,E^{v}\le\,0$, without the strong hybridization with the C-dominated $\pi$ bands.
They have the almost flat dispersions, mainly owing to the very weak chemical bondings between (Cl,Br) and carbon atoms.
Specifically, the Cl and Br atoms can induce the ferromagnetic spin configurations. Their significant splitting of the spin-up and spin-down energy bands is revealed near $E_{F}$, and the largest energy spacing is $\sim 0.5$ eV.

The main features of energy bands can be verified by ARPES measurements. As for few-layer graphenes, the diverse electronic properties have been identified by ARPES, especially for the sensitive changes of energy bands with the number of layers,\cite{sutter2009electronic,ohta2007interlayer} stacking configurations,\cite{ohta2006controlling} gate voltages,\cite{zhou2007substrate} and adatom absorptions.\cite{zhou2008metal,schulte2013bandgap} The high-resolution ARPES observations on the Li-, Na-, and K-doped graphenes have confirmed many free electrons in the linear conduction band.\cite{virojanadara2010epitaxial,sugawara2011fabrication,papagno2011large,gruneis2008tunable} These systems reveal the red shift of $\sim 1.3$ eV of the the $\pi$ and $\pi^{*}$ bands, being in good agreement with our calculations. On the other hand, the same method is performed on F- and Cl-intercalated gaphenes,\cite{walter2011highly,vinogradov2012controllable} revealing blue shifts of low-lying bands.
ARPES can be further utilized to examine the feature-rich band structures of Al- and halogen-doped graphenes, including the shift or destruction of the Dirac cone, the adatom-dominated valence and conduction bands, the spin-dependent energy bands, and the diverse energy dispersions near E$_F$.

The carrier density ($\rho$) and the variation of carrier density ($\Delta \rho$) can provide very useful information in the orbital bondings and energy bands.\cite{lin2015feature,lin2015h}
The former directly reveals the bonding strength of C-C, Al-C, alkali-C, and halogen-C bonds, as illustrated in Figs. 4(a)-4(e).
As to the Al- and Li-doped graphenes, all the C-C bonds possess the strong covalent $\sigma$ bonds and the weak $\pi$ bonds simultaneously (black and grey rectangles in Figs. 4(b)-4(d)).
Such bondings are only slightly changed under various adatom concentrations, as shown by Al-doped systems in Figs. 4(b) and 4(c). This is responsible the quasi-rigid shifts of the $\pi$ and $\sigma$ bands.
On the other hand, the F-doped graphene (Fig. 4(e)) shows a strong covalent F-C bond, a weakened $\sigma$ C-C bond (black rectangle), and the slightly deformed $\pi$ bonds (grey rectangle). Apparently, these can create the deep F-dominated bands, the more complicated $\pi$ and $\sigma$ bands, and the seriously distorted Dirac cone.

\begin{figure}[h]
\centering
  \includegraphics[width=1.0\linewidth]{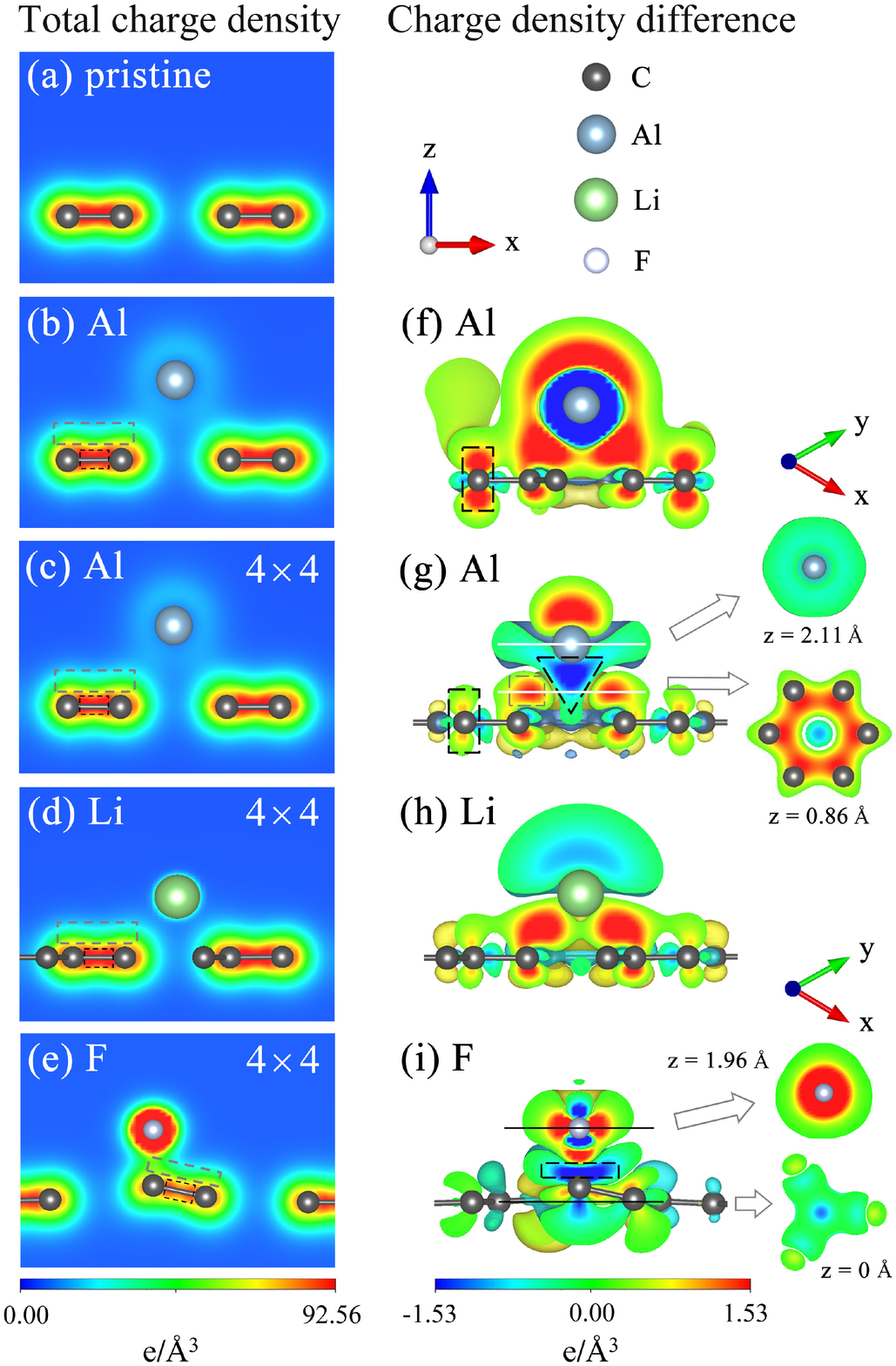}
  \caption{The spatial charge density for (a) pristine graphene, (b) Al-doped graphene in a $2\times\,2$ cell; (c) Al- (d) Li- and (e) F-doped graphenes in a $4\times\,4$ cell. Similar plots corresponding to the density differences are shown in (f), (g), (h); (i) for adatom-doped graphenes. Also revealed in the insets are projections on xy plane.}
  \label{fgr:4}
\end{figure}

The charge density difference is very useful in understanding the electron transfer and orbital bondings, as illustrated in Figs. 4(f)-4(i). $\Delta \rho$ is created by subtracting the carrier density of pristine graphene from that of adatom-doped graphene.
For Al-doped systems (Figs. 4(f) and 4(g)), $3s$-orbital electrons (dark blue; triangle) are redistributed between Al and the six nearest C atoms, revealing a significant hybridization of $3s$ and $2p_z$ orbitals (grey rectangle; red ring in the inset). The spatial charge distribution of ($3p_x$,$3p_y$) orbitals is extended from the light blue ring (inset) near Al to the red ring between C and Al atoms. It is noticed that these orbitals make less contribution to the chemical bonding. The multi-orbital hybridizations of $3s$ and ($3p_x$,$3p_y$) are, respectively, associated with the Al-dominated valence and conduction bands (Fig. 2(d)). The electrons of Al adatom are transferred to the top and bottom of the non-nearest C atoms (red region within a black rectangle), leading to free carriers in conduction Dirac cone. The higher the Al-concentration is, the more the electrons are transferred (Fig. 4(f)). Also, the Li-doped graphene exhibits the similar orbital hybridization of $2s$ and $2p_z$ orbitals (Fig. 4(h)). The $\pi$-bondings are almost identical in the Al- and alkali-doped systems, and so do the preserved Dirac-cone structures. It should be noticed that the Li-dominated conduction band arises from the hybridization of $2s$ and $2p_z$ orbitals (grey rectangle), and the Li-dominated valence bands are absent.
As to the F-doped graphene (Fig. 4(i)), there exist very complicated charge density differences in C-C and F-C bonds, owing to the rather strong interactions between adatoms and carbons. The $sp^3$ bonding is revealed in a buckled graphene structure, as indicated from the deformations of the $\sigma$ bonding in the nearest C and the $\pi$ bonding in the next-nearest C. Charges are transferred from C to F. The obvious spatial distribution variations on xz and xy planes clearly illustrate that the multi-orbital hybridizations of ($2p_x,2p_y,2p_z$) in F-C bonds. These are closely related to the largest electronegativity of F.

The main characteristics of DOS are determined by the band-edge states of energy dispersions; furthermore, the orbital-projected DOS directly reflects the feature-rich chemical bondings. The Al-doped graphene exhibits a lot of 2D Van Hove singularities, as indicated in Figs. 5(a)-5(f).
A dip, strong symmetric peaks, and shoulder structures, respectively, come from the linear bands, the saddle point, and the local extreme points.
Graphene possesses the $2p_z$-dominated DOS in the range of $-3$ eV $\le\,E\le\, 2$ eV (the red curve in Fig. 5(a)), including the dip at $E=0$, the linear energy E-dependence, and the ($\pi$,$\pi^{*}$) peaks in the logarithmically divergent form at ($-2.4$ eV,$1.7$ eV).
The special structures in the $2p_z$-related DOS are dramatically changed by the orbital hybridization of $3s$ and $2p_z$ (Fig. 5(b)-5(f)).
DOS due to the conduction $\pi^{*}$-states is finite at the Fermi level, being a result of high free carrier density. The center of the linear E-dependence, which depends on the adatom concentrations, is changed from zero energy to the range of $-1.5$ eV$\le\,E\le\,-0.8$ eV. The prominent $\pi$ peak becomes several sub-peaks in the range of $-5$ eV $\le\,E\le\,-2.5$ eV, because of the zone-folding effect. Part of them are merged with the Al-created peaks at $-3.5$ eV $\le\,E\le\,-2.5$ eV (the dashed blue curve), clearly illustrating the significant hybridization of $3s$ and $2p_z$ orbitals. In addition, there are only weak peaks related to the adatom $3p_z$ orbitals (the dashed pink curve). Moreover, the low-lying splitting $\pi^{*}$ peaks are distributed in $0,\le\,E\le\,-1.5$ eV, especially for certain peaks very close to the Fermi level ($2 \times 2$ case in Fig. 5(b)). Two strong peaks arising from the Al ($3p_x$,$3p_y$) orbitals are also revealed in a narrower energy range (the dashed green curve). On the other hand, the $\sigma$-band shoulder structure due to $2p_x$ and $2p_y$ orbitals exhibits a red shift of $E=-3$ eV $\rightarrow$ ~$-4$ eV after the aluminum adsorption, regardless of Al-concentrations (green curves in Figs. 5(b)-5(f)).

\begin{figure}[h]
\centering
  \includegraphics[width=1.0\linewidth]{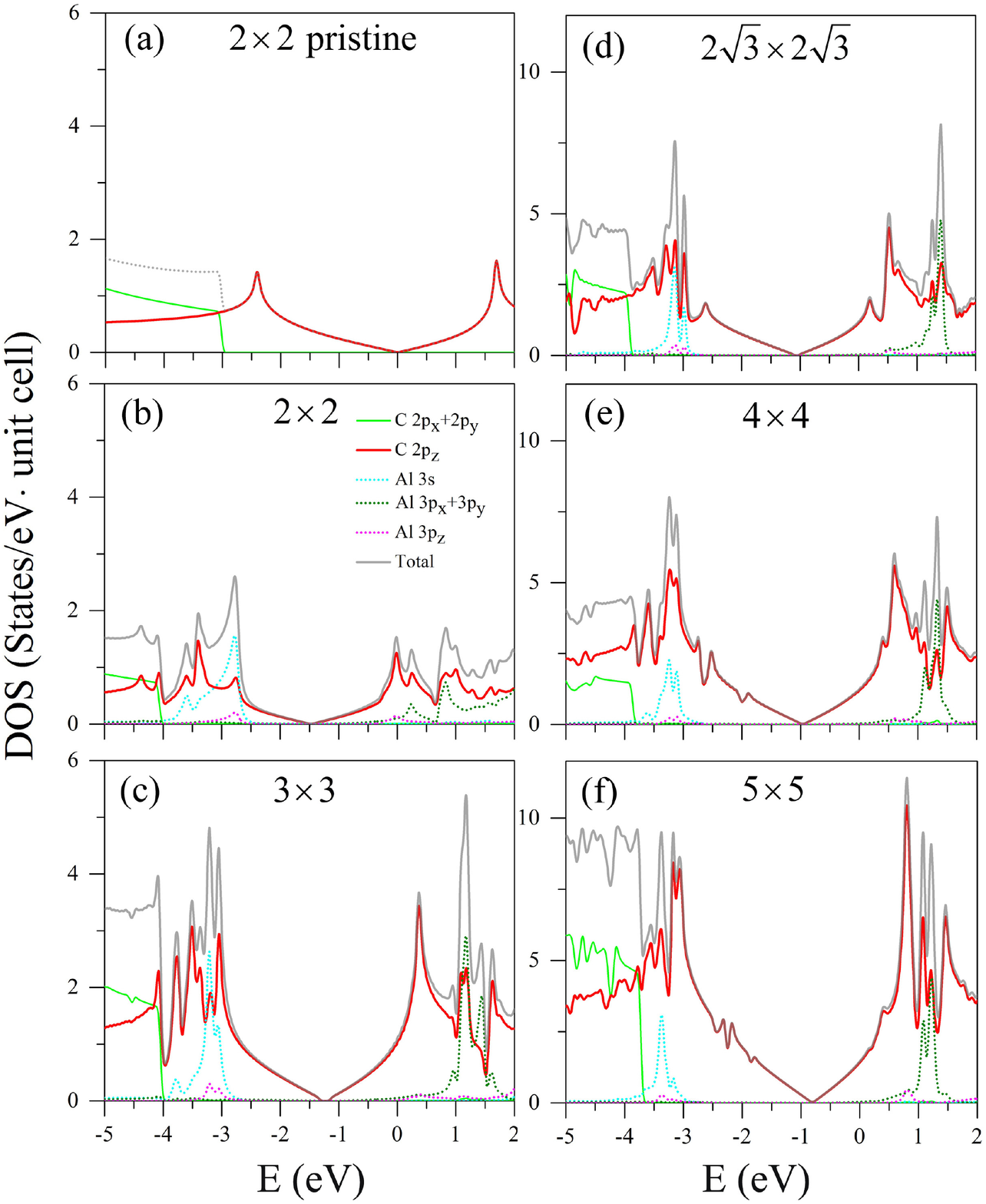}
  \caption{The total and the orbital-projected DOSs for Al-doped graphenes with distinct  adsorption cases: (a) pristine, (b) $2\times\,2,$ (c) $3\times\,3,$ (d) $2\sqrt 3\times\,2 \sqrt 3\,$ (e) $4\times\,4,$  and (f) $5\times\,5$}
  \label{fgr:5}
\end{figure}

The special chemical bondings cause the alkali-doped graphenes to exhibit the similar DOS structures, as shown in Figs. 6(a)-6(c). For Li-, Na-, and K-doped graphenes, a low-lying symmetric peak is revealed at $E=1.0$ eV, $0.1$ eV, and $0.4$ eV, respectively. This peak, which directly reflects the hybridizations of $2p_z$ and the outmost $s$ orbitals, mainly comes from the alkali-dominated conduction bands with the weak energy dispersions.
The DOS at the Fermi level is comparable for the Li-, K-, and Al-doped systems except for the Na-doped graphene, since the enhanced DOS in the last system is contributed by Na adatoms.
Moreover, the similar structures in DOS between alkali- and Al-doped graphenes include the dip, strong symmetric peaks, and shoulder structures. Specifically, these systems have the almost identical DOS in $-2$ eV$\le\,E\le\,0$. On the other hand, the merged peaks due to the splitting $\pi$ peaks and the alkali-created peaks are absent in $-5$ eV $\le\,E\le\,-2.5$ eV.
The main difference between alkali- and Al-doped graphenes lies in the much significant orbital hybridization in the latter, as indicated by the Al-dominated valence bands.

\begin{figure}[h]
\centering
  \includegraphics[width=1.0\linewidth]{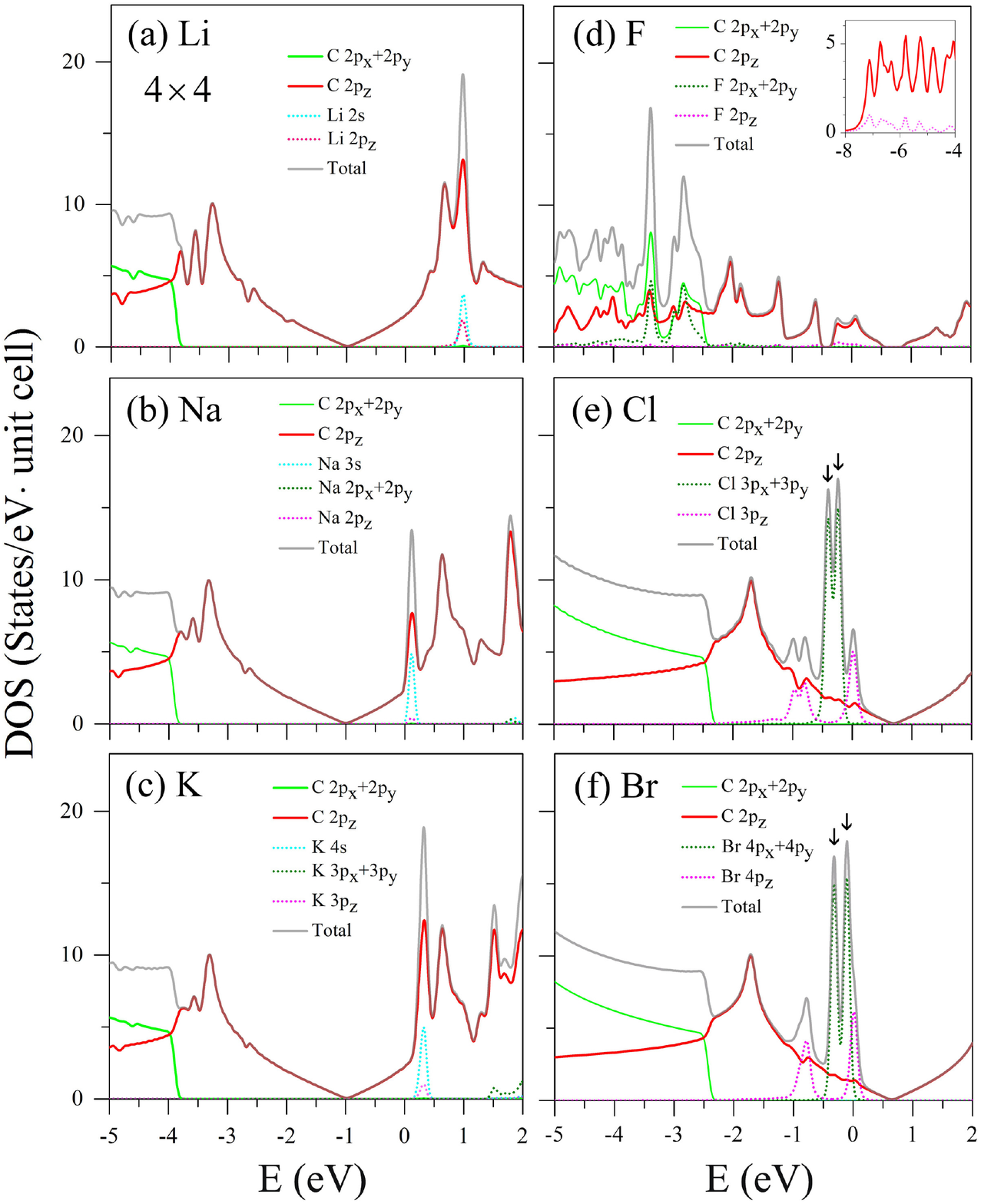}
  \caption{Same plots as Fig. 5, but shown for (a) Li-, (b) Na-, (c) K-, (d) F-, (e) Cl-, and (f) Br-doped graphenes in a $4\times\,4$ cell. Inset in (d) is DOS at the deeper energy.}
  \label{fgr:6}
\end{figure}

In sharp contrast to the Al- and alkali-doped graphenes, halogen-doped graphenes have the more complicated special structures in DOS. Such structures strongly depend on the strength of chemical bondings between halogen and carbon atoms. The F-doped graphene exhibits a lot strong peaks related to the $2p_z$ orbitals of C atoms, as shown in Fig. 6(d). Such structures are merged with the ($2p_x$,$2p_y$) peaks in $-5$ eV $\le\,E\le\, -2.5$ eV, reflecting the significant $sp^3$ C-C bonding in the buckled graphene structure (Figs. 4(e) and 4(i)). Furthermore, the ($2p_x$,$2p_y$,$2p_z$) peaks of C atoms and the ($2p_x$,$2p_y$) peaks of F atoms are revealed at $-2.5$ eV $\le\,E\le\,-3.5$ eV simultaneously. In addition, many $2p_z$- subpeaks of F atoms
are revealed at deeper energy (the dashed pink curve in inset). All the peaks mainly come from the complicated orbital hybridizations in F-C and C-C bonds. The finite DOS at the Fermi level is induced by the $\pi$ bondings of carbon $2p_z$ orbitals, being associated with the valence Dirac cone. On the other hand, the Cl- and Br-doped graphenes, as shown in Figs. 6(e) and 6(f), have a dip structure at E$\sim\,0.7$ eV; that is, there exist the blue-shift Dirac structures.
They exhibit only few $2p_z$-orbtal peaks due to carbon atoms; furthermore, the strong $\pi$ peak and the $\sigma$ shoulder at E$\le\,-1.5$ eV are similar to those of pristine graphene (Fig. 5(a)). Specifically, the adatoms can create some strong peaks related to ($3p_x$,$3p_y$,$3p_z$) or ($4p_x$,$4p_y$,$4p_z$) orbitals. Such peaks are slightly combined with the weak $2p_z$-orbital peaks. Furthermore, they correspond to the splitting spin configurations (the up and down states shown by the black arrows). In addition, the energy spacings ($>0.2$ eV) between two splitting peaks are sufficiently wide for the experimental examinations. The above-mentions features clearly present the strength of the orbital hybridizations in halogen-carbon bonds.

The form, energy, number and intensity of special structures in DOS can be examined by the STS measurements. The tunneling conductance (dI/dV) is approximately proportional to DOS and directly reflects the main features in DOS. This powerful method has been successfully utilized to investigate the diverse electronic properties in the carbon-related systems, such as, carbon nanotubes,\cite{wilder1998electronic,chen2013tuning} graphites [Rs], graphene nanoribbons,\cite{huang2012spatially,sode2015electronic} few-layer graphenes,\cite{lauffer2008atomic,choi2010atomic} and adatom-adsorbed graphenes.\cite{tapaszto2008tuning,gyamfi2011fe} For example, the
band gap of exfoliated oxidized graphene sheets\cite{pandey2008scanning} and the Fermi-level red shift of Bi adatoms on graphene surface\cite{chen2015long,chen2015tailoring} are confirmed by STS.
The predicted DOSs in Al-, alkali- and halogen-doped graphenes, which include the dip structure, the shift of the Fermi level, the splitting or preserved $\pi$- and $\pi^{*}$-peaks, the $\sigma$ shoulder, the adatom-created peaks, and the spin-split low-energy peaks, could be further verified with STS. The STS measurements are useful in comprehending the specific orbital hybridizations and the distinct doping effects.

\section{CONCLUDING REMARKS}

The geometric and electronic properties of Al-doped graphenes are studied using the first-principles calculations. They are shown to be dominated by the diverse orbital hybridizations in chemical bonds. The $\pi$- and $\sigma$-bondings of carbon atoms are slightly changed after adatom adsorption, while charges are transferred from Al to C. Also, there exist the significant hybridizations between ($3s$,$3p_x$,$3p_y$) and $2p_z$ orbitals in the Al-C bonds. These are responsible for the quasi-rigid shifts of energy bands, the free electrons in the conduction Dirac cone, and the Al-dominated valence and conduction bands. Specially, the Al-induced high free carrier density is almost same with that of the alkali doping. Such systems might have high potentials for future technological applications, e.g., high-capacity batteries,\cite{lin2015ultrafast,rani2013fluorinated} and energy storage.\cite{ao2009doped,ao2010high} The rich features of energy bands lead to many special structures in DOS, such as, a dip with the linear E-dependence, strong peaks and shoulder structures. The experimental examinations using ARPES and STS can provide the full information about the critical orbital hybridizations.

The essential properties are diversified by the distinct adatom absorptions. The alkali- and Al-doped graphenes have the almost identical red-shift Dirac-cone structure, reflecting the slight change of the $\pi$ bonding. However, the adatom-created valence bands are absent in the alkali-doped systems because of a single electron in the outmost $s$ orbital. As a result of the only $s$-$2p_z$ hybridization in alkali-C bonds, the former present simple $\pi$-peak structures, and several strong $\pi^{*}$ peaks accompanied with a alkali-induced peak. On the other hand, the halogen-doped graphenes exhibit free holes in the valence Dirac cone. Specifically, the quite strong F-C bondings mainly come from the very complicated ($2p_x,2p_y,2p_z$) orbital hybridizations. They result in the lowest absorption energy, the seriously distorted Dirac cone, the buckled graphene structure (or the $sp^3$ bonding), and the more special structures in DOS. As to the Cl- and Br-doped graphenes, there exist very weak hybridizations, respectively, due to the ($3p_x,3p_y,3p_z$)-$2p_z$ orbitals and ($4p_x,4p_y,4p_z$)-$2p_z$ orbitals.
This causes the adatom-dominated bands to appear near the Fermi level and exhibit the rather prominent peaks in the low-energy DOS. Furthermore, these two systems possess the ferromagnetic spin configuration responsible for the splitting low-lying energy bands and DOS peaks. In short, the feature-rich electronic properties of the adatom-doped graphenes can be created by the diverse orbital hybridizations in the chemical bonds.

\vskip 0.6 truecm
\par\noindent {\bf ACKNOWLEDGMENTS}
\vskip 0.3 truecm

This work was supported by the Physics Division, National Center for Theoretical Sciences (South), the Nation Science Council of Taiwan (Grant No. NSC 102-2112-M-006-007-MY3). We also thank the National Center for High-performance Computing (NCHC) for computer facilities.

\footnotesize{
\bibliography{article_Xavier} 
\bibliographystyle{rsc} 
}


\end{document}